\documentclass[12pt,preprint]{aastex}

\slugcomment{}

\shorttitle{3D Coronal Density and the Magnetic Field during Solar Minimum}
\shortauthors{M. Kramar {\it et al.}}

\usepackage{graphicx}                    
\usepackage{float}
\usepackage{color}                       
\usepackage{url}                         
\usepackage{hyperref}
\usepackage{ulem}

\usepackage{enumitem}
\setlist{nolistsep}

\usepackage{tikz}

\usepackage{bm}

\begin{document}

\title{3D Coronal Density Reconstruction and Retrieving the Magnetic Field Structure during Solar Minimum}




\author{M. Kramar\altaffilmark{1,2}, V. Airapetian\altaffilmark{2}, Z. Miki\'{c}\altaffilmark{3}, J. Davila\altaffilmark{2}}

\altaffiltext{1}{The Catholic University of America, 620 Michigan Ave NE, Washington, DC 20064, USA}
\altaffiltext{2}{NASA-GSFC, Code 671, Greenbelt, MD 20771, USA}
\altaffiltext{3}{Predictive Science, Inc., 9990 Mesa Rim Rd., Ste. 170, San~Diego, CA 92121, USA}

\begin{abstract}
Measurement of the coronal magnetic field is a crucial ingredient in understanding the nature of solar coronal phenomena at all scales. 
We employed STEREO/COR1 data obtained during a deep minimum of solar activity in February 2008 (Carrington rotation CR 2066) to 
retrieve and analyze the three-dimensional (3D) coronal electron density in the range of heights 
from $1.5$ to $4\ \mathrm{R}_\odot$ using a tomography method. 
With this, we qualitatively deduced structures of the coronal magnetic field. 
The 3D electron density analysis is complemented by the 3D STEREO/EUVI emissivity in the 195 \AA \ band obtained 
by tomography for the same CR. 
A global 3D MHD model of the solar corona was used to relate the reconstructed 3D density and emissivity to 
open/closed magnetic field structures. 
We show that the density maximum locations can serve as an indicator of current sheet position, 
while the locations of the density gradient maximum can be a reliable indicator of coronal hole boundaries. 
We find that the magnetic field configuration during CR 2066 has a tendency to become radially open 
at heliocentric distances greater than $2.5 \ \mathrm{R}_\odot$. 
We also find that the potential field model with a fixed source surface (PFSS) is inconsistent 
with the boundaries between the regions with open and closed magnetic field structures. 
This indicates that the assumption of the potential nature of the coronal global magnetic field is 
not satisfied even during the deep solar minimum. 
Results of our 3D density reconstruction will help to constrain solar coronal field models and test the accuracy of 
the magnetic field approximations for coronal modeling. 
\end{abstract}


\section{Introduction}

Solar coronal magnetic fields play a key role in the energetics and dynamics of coronal heating, solar flares, coronal mass ejections, 
filament eruptions, and determine space weather processes. 
Therefore, one of the central problems of solar physics is to measure the magnetic fields in the solar corona. 
However, currently available routine extrapolation methods do not provide direct ways for 
characterizing global magnetic fields in the solar corona. 

The main techniques that are currently used to deduce the global magnetic structure of the solar corona 
include potential field source surface (PFSS) models, 
nonlinear force-free field (NLFFF) models, and multidimensional magnetohydrodynamic (MHD) models of the global solar corona.
These methods are based on boundary conditions of the solar photospheric magnetic 
field that are derived directly from photospheric magnetograms. 
The PFSS model is a relatively simple model, which is routinely used to extrapolate 
the photospheric magnetic field into the global solar corona 
\citep{Altschuler_1969,Schatten_1969SoPh,Wang_Sheeley_1992,Luhmann_2002,Schrijver_2003}. 
It assumes that the magnetic field is current-free between the photosphere (the inner boundary) and the source surface (outer boundary). 
Its inner boundary is based on synoptic maps of photospheric magnetograms. 
The outer boundary represents a spherical source surface with constant radius, 
typically ranging from $1.5$ to $3.5 \ \mathrm{R}_\odot$ \citep{Lee_Luhmann_2011}. 
Since the potential magnetic field is the field with the lowest minimum energy for 
a given photospheric radial boundary condition \citep{Sakurai_1989}, it cannot 
account for dynamical processes such as eruptions, flares, and magnetic reconnection, 
during which magnetic energy is converted into plasma kinetic energy, 
without significantly changing the magnetic boundary flux.
Indeed, soft X-ray observations of active regions often show a non-potential structure of the magnetic field 
\citep{Jiao_1997}.

The NLFFF model is a more advanced step in extrapolating the surface magnetic field into the corona. 
It is suitable for use with recently available data from the {\it Helioseismic and Magnetic Imager} (HMI) 
onboard the {\it Solar Dynamic Observatory} (SDO) 
and {\it Vector SpectroMagnetograph} (VSM) at the National Solar Observatory (NSO). 
Unlike the PFSS model, it assumes that the current is parallel to the magnetic field. 
This approach uses the photospheric vector magnetograms, such as those from
the HMI and VSM instruments to extrapolate the surface data into the solar corona 
\citep{Wiegelmann_2005, Wiegelmann_2008, Tadesse_2014AA}, 
and therefore provides a better description of the coronal magnetic field. 
However, the NLFFF method is not suitable for determining the magnetic field 
if the force-free assumption is not satisfied everywhere in the volume of the extrapolation 
\citep{Demoulin_1992, Gary_2001}. 
In addition, this model does not provide information about the plasma density or temperature of coronal structures, and therefore
cannot be used to predict emission measure, so the results cannot be compared with extreme ultraviolet (EUV) observations.

Complementary to the PFSS and NLFFF extrapolation methods, a number of self-consistent magnetohydrodynamic (MHD) 
models of the solar corona have been developed 
\citep{Mikic_1999, Riley_2001, Mikic_2007, Lionello_2009, Airapetian_2011, Toth_2012, van_der_Holst_2014}. 
Unlike the PFSS or NLFFF techniques, this approach includes a self-consistent time-dependent treatment 
of the plasma pressure, gravitational and 
magnetic forces that are required to describe the dynamics of helmet streamers, coronal mass ejections, and the solar wind. 
However, the application of these models is limited by approximations used for describing the coronal heating, 
and the uncertainties in the boundary conditions that are deduced from synoptic data. 
Therefore, these complex models need to be validated by direct observations of the coronal magnetic field.

All of these methods are essentially extrapolation methods based on inner boundary conditions taken at the photosphere. 
However, the magnetic field at the photosphere and the lower chromosphere is far from 
potential or force-free, because of the dominance of the plasma pressure there.
It has been suggested that chromospheric magnetograms are better suited as boundary conditions 
for extrapolation methods \citep{Judge_2010}. 
Sophisticated multidimensional MHD--RHD (radiation hydrodynamics) models of the 
solar chromosphere are currently under development.

Direct measurements of the coronal magnetic field are among the most challenging problems in observational solar astronomy. 
Significant progress has recently been achieved here with the deployment of the {\it Coronal Multichannel Polarimeter} (CoMP) 
of the High Altitude Observatory (HAO). 
The instrument provides polarization measurements of the Fe~{\sc{xiii}} 10747 \AA \  forbidden line emission 
\citep{Tomczyk_2007, Tomczyk_2009}. 
The observed polarization depends on the magnetic field through the coronal Hanle and Zeeman effects 
\citep{Charvin_1965, Sahal_1977, House_1977, Casini_1999, Lin_2000}. 
To use this type of data, 
the vector tomography method has been developed for 3D reconstruction of the coronal magnetic field 
\citep{Kramar_2006, Kramar_2013}. 
However, because of the small field of view (FOV) of the CoMP instrument, 
it would be problematic to reliably reconstruct the coronal magnetic field 
above $\approx 1.2 \ \mathrm{R}_\odot$ based on such CoMP observations \citep{Kramar_2013}. 
In this respect, the {\it Solar Terrestrial Relations Observatory} (STEREO) COR1 
coronal observations provide a unique opportunity to characterize 
global coronal conditions at heights greater than $\approx 1.4 \ \mathrm{R}_\odot$.

In this article, we study the 3D structure of coronal streamers to determine the height at which 
the coronal magnetic field becomes radial. 
Specifically, we use data from the STEREO/COR1 coronagraph for half a solar rotation period 
during CR 2066 to reconstruct the 3D coronal electron density with the tomography method. 
Our results are complemented by the 3D emissivity obtained by tomography for 
the STEREO/{\it Extreme Ultraviolet Imager} (EUVI) data in the 195 \AA \ band. 
We tested the tomography method for systematic errors with simulated pB-data produced by integrating 
the results of a 3D thermodynamic MHD model over the line of sight (LOS). 
Finally, we compare the reconstructed 3D coronal structures with the PFSS model.

\section{Tomography}
\label{Sect_Tomo}

For wavelengths for which the corona is optically thin, the radiation coming from the corona is a
LOS integral of the emissivity in the observed direction. 
Therefore, it is impossible to reconstruct the spatial distribution of the
emissivity from a single (in a geometric sense) measurement or projection.
The solution space is reduced if we have measurements from many different viewpoints. 
The reconstruction based on the observations of an object from different view angles is essential for tomography. 
The possibility of reconstructing a function from its projections was first studied by \cite{Radon_1917}. 
Several decades later, this purely mathematical research formed the basis for 
the tomography method, which was developed to reconstruct the X-ray absorption coefficient in human bodies.
The first experimental X-ray tomographic scanner was made by \cite{Hounsfield_1972}, 
and \cite{Cormack_1963, Cormack_1964} independently discovered some of the algorithms for the reconstruction. 
These two authors received the Nobel prize for their investigations in 1979. 
Today, tomography is used in many fields: medicine, material structure testing, geophysics, 
astrophysics \citep{Boffin_2001}. 
In solar coronal physics, the use of tomography was first proposed by \cite{Wilson_1976} and later by \cite{Davila_1994}. 
In astrophysical applications the input data can suffer from noise and data incompleteness. 
However, the regularization method allows solar coronal tomography to produce reliable reconstructions 
\citep{Tikhonov_1963, Frazin_2002, Kramar_2006, Kramar_2009}  
(see also Section \ref{Sect_Test_Tomo_COR1} of this article).

\subsection{Tomography Based on White-Light STEREO/COR1 Data}
\label{Sect_Tomo_COR1}

To reconstruct extended coronal structures, the reconstruction algorithm requires 
observations from more than two directions. 
This is the key requirement of tomography. 
Tomography applications for coronal studies typically assume a rigid rotation of the coronal density structures. 
The algorithm requires coronagraph data for half a solar rotation as input if observed from a single spacecraft, and, generally, 
coronal structures that are stable over their observation periods can reliably be reconstructed 
\citep{Davila_1994, Zidowitz_1999, Frazin_2005, Kramar_2009}. 
However, depending on the positions of a coronal structure relative to the spacecraft during the observation period, 
the stationarity assumption for that structure can be reduced to about a week \citep{Kramar_2011}. 

For our density reconstructions we used the polarized brightness (pB) intensity images from the COR1 instrument 
onboard the STEREO-B spacecraft 
taken 28 images per half a solar rotation as input for the tomographic inversion. 
We limited here the data input for the tomography based on COR1 data to the STEREO-B spacecraft 
because COR1-B had lower levels of stray light during CR 2066 than COR1-A. 

In the STEREO/COR1-B field of view (below $\approx 4\ \mathrm{R}_\odot$), the white-light pB coronal emission is dominated by 
scattering sunlight on the free electron in the corona 
\citep{Blackwell_1966a, Blackwell_1966b, Moran_2006, Frazin_2007}. 
The intensity of the pB-signal as a fraction of the mean solar brightness is given as 
\begin{equation}
I_\textrm{\small{pB}}(\vec{\hat{e}}_\textrm{\scriptsize{LOS}},\vec{\rho})=
\int \limits_\textrm{LOS} K(\vec{r})  N_e(\vec{r}) \mathrm{d} \ell ,
\label{ThompsonScat_pB}
\end{equation}
where $N_e$ is the electron density, $\vec{\rho}$ is a vector in plane-of-sky (POS) from the Sun center to the LOS 
and perpendicular to LOS, $\ell$ is length along the LOS, 
$\vec{\hat{e}}_\textrm{\scriptsize{LOS}}$ is the unit vector along the LOS, 
and $\vec{r}$ is the radius-vector. 
The kernel function [$K$] is defined by the Thompson scattering effect 
\citep{van_de_Hulst_1950, Billings_1966, Quemerais_2002}: 
\begin{equation}
K=\frac{\pi \sigma}{2\left (1-\frac{u}{3} \right )}
\left [ (1-u)A(r)+uB(r) \right ] \frac{\rho^2}{r^2}
\label{KernelFunction_pB}
\end{equation}
where the expressions for $A(r)$ and $B(r)$ are the same as those given by \cite{Quemerais_2002}, 
$\sigma=7.95\times 10^{-26}\ \textrm{cm}^2$ is the Thompson scattering cross-section for a single electron, 
and the linear limb-darkening coefficient [$u$] is set to $0.6$ in the present calculations.

Because COR1 views the corona close to the limb, the instrument has a significant amount of scattered light, 
which must be subtracted from the image 
prior to be applied in the reconstruction method. 
Proper removal of instrumental scattered light is essential for coronal reconstruction.  
One way is to subtract a monthly minimum (MM) background. 
The monthly minimum approximates the instrumental scatter by finding the lowest value of each pixel in all images 
during a period of about one month. 
However, this method tends to overestimate the scattered light in the streamer belt (equatorial region). 
The lowest value of these pixels during a month will contain both the scattered light and the steady-intensity value from the corona. 
Hence, if we were to use such pixels as input for our electron density reconstruction, 
we would obtain an electron density that is lower than the actual density. 

Another way to account for the scattered light is to subtract a roll minimum (RM) background. 
The roll minimum background is the lowest value of each pixel obtained during a roll maneuver of the spacecraft (instrument) around its optical axis. 
Because the coronal polar regions are much darker than the equatorial ones, the lowest pixel values in the equatorial region 
during the roll maneuver are nearer to the value of the scattered-light intensity than the MM. 

The sensitivity of the COR1-B instrument decreases at a rate of about $0.25$~\% per month \citep{Thompson_2008}. 
Moreover, variations in the spacecraft's distance from the Sun cause changes of the amount of scattered light in the coronagraph images. 
But the roll maneuvers are done rarely. 
Therefore it is impossible to use an RM background obtained in one month for data from another month 
when the highest possible photometric accuracy is needed. 
One ways to obtain a background image for the period between the roll maneuveres 
is to interpolate RM backgrounds over time in a such a way that this temporal dependence follows 
the temporal dependence of the MM backgrounds, because the MM background images are available for every month. 
This approach is realized by W. Thompson in the SolarSoft IDL routine {\sf secchi\_prep} with the keyword parameter {\sf calroll}. 
We used backgrounds obtained in this way. 
The photometric calibration is based on Jupiter's passage through the COR1 FOV  \citep{Thompson_2008}. 

After subtracting the scattered light, a median filter with a width of three pixels was applied 
to reduce anomalously bright pixels caused by cosmic rays. 
Then, every third image pixel was taken (resulting in a $340\times 340$ pixel image) to reduce the computer memory size. 
The reconstruction domain is a spherical grid with a size of $50\times180\times360$ covering 
heliocentric distances from $1.5$ to $4\ \mathrm{R}_\odot$, Carrington latitudes from $-90$ to $90^\circ$, 
and Carrington longitudes from $0$ to $360^\circ$, respectively.

The inversion was performed for the function 
\begin{equation}
F=\left | \mathbf{A}\cdot\mathbf{X}-\mathbf{Y} \right |^2 + \mu \left | \mathbf{R}\cdot\mathbf{X} \right |^2. 
\label{MinFun}
\end{equation}
Here, the elements $x_j$ of the column matrix $\mathbf{X}$ contain the values of electron density [$N_e$] in the grid cells 
with index $j=1,...,n$, 
and $y_i$ is the data value for the $i$-th ray, 
where index $i=1,...,m$ accounts for both the viewing direction and pixel position in the image.  
The element $a_{ij}$ of the matrix $\mathbf{A}$ represents the intersection of 
volume element $j$ with LOS related to pixel $i$, multiplied by the kernel function 
that is defined by the Thompson scattering effect for the pB-intensity signal (see Equation (\ref{ThompsonScat_pB})). 
The second term on the right-hand side of Equation (\ref{MinFun}) is the 
regularization term that minimizes the effects of noise and data gaps \citep{Tikhonov_1963}. 
The matrix $\mathbf{R}$ is a diagonal-like matrix such that the regularization is the first-order smoothing term, 
{\it i.e.} operation $|\mathbf{R}\cdot\mathbf{X}|^2$ produces 
the square difference in value between two neighboring grid cells, summed over all cells. 
The regularization parameter [$\mu$] regulates balance between the smoothness of the solution on one hand 
and the noise and reconstruction artifacts on the other.  
The result of the inversion depends on a number of factors, including the number of iterations and the value of $\mu$.
The value of $\mu$ was chosen using the cross-validation method \citep{Frazin_2002}.  
The iterations are performed until the first term in Equation (\ref{MinFun}) became slightly lower than the data noise level, 
which is essentially the Poisson noise in the data. 

The coronal electron density drops very rapidly with distance from the Sun, introducing a wide dynamic range in the data, 
which causes linear artifacts in the reconstruction. 
To increase the contribution of signals from those LOS that pass through the low density regions, 
and to reduce the artifacts in the numerical reconstruction at larger distances from the Sun, 
we applied a set of weighting coefficients (or preconditioning) 
\begin{equation}
w_i =
\frac{1}{\left( y_i^{(FT1)} \right)^2}
\label{WeightingFunGeneral}
\end{equation}
were applied for the first term in Equation (\ref{MinFun}) in such way that $\sum\limits_{j} (w_i a_{i,j} x_j) = w_i y_i $. 
Here, $y_i^{(FT1)}$ is the inverse Fourier transform of the function $y_i(r_p,\phi_p)$ on $\phi_p$
with harmonics taken up to first order, where $y_i(r_p,\phi_p)$ is the data value at the position $(r_p,\phi_p)$ 
in the polar coordinate system for some particular image.  
The value of $r_p$ is fixed for a given pixel and set equal to the radial distance from the center of the Sun's disk to the pixel.  
A more detailed description of used tomography method is given in \cite{Kramar_2009}. 
The error estimation of the tomographic method is investigated below in Section \ref{Sect_Test_Tomo_COR1}. 
The reconstruction results are discussed in Section \ref{Sect_CR2066}.

\subsection{Tomography for Emissivity from STEREO/EUVI Data}
\label{Sect_EUVI_Tomo}

The STEREO/EUVI instrument observes the corona up to about $1.7 \ \mathrm{R}_\odot$ 
in four spectral channels (171, 195, 284, and 304 \AA) that span the 0.1 to 20 MK temperature range 
\citep{Wuelser_2004, Howard_2008}. 
The measured coronal emission in the 171, 195, and 284 \AA \  channels can be represented as the result of emission integrated over the LOS as 
\begin{equation}
I(\hat{\vec{e}}_\textrm{LOS},\vec{\rho})=
k  \int \limits_\textrm{LOS} \varepsilon(\vec{r}) \mathrm{d} \ell ,
\label{EUVI_Emiss_LOS}
\end{equation}
where $\varepsilon(\vec{r})$ is the emissivity at the position $\vec{r}$ in the selected channel, 
{\it i.e.} light intensity (in photons per second for example) emitted per unit volume, per unit solid angle. 
The coefficient $k$ accounts for pixel size, aperture, and distance to the Sun. 

As input, we used EUVI 195 \AA \  images calibrated by applying IDL SolarSoft routines. 
To reduce anomalously bright pixels caused by cosmic rays, the IDL SolarSoft routine {\sf despike\_gen} was applied. 
Three images taken with about two hours difference were averaged into one. 
Three averaged images per day were taken during a period of half a solar rotation. 
Then, every fourth pixel was taken, resulting in $512\times512$ input image. 

We inverted $\varepsilon(\vec{r})$ in the same manner as for the electron density 
in the white-light tomography with $K$ and $N_e$ in Equation (\ref{ThompsonScat_pB}) substituted by $k$ and $\varepsilon$, respectively, 
according to Equation (\ref{EUVI_Emiss_LOS}). 
The inversion result is 
the 3D emissivity distribution for the EUVI 195 \AA \  channel in the coronal range from $1.05$ to $1.5\ \mathrm{R}_\odot$. 
Figure \ref{Fig_CR2066_EUVI_Rec_sph} shows a spherical cross-section of the reconstructed EUVI 195 \AA \  emissivity 
at a heliocentric distance of $1.1 \ \mathrm{R}_\odot$ for CR 2066. 
The reconstruction result is discussed in Section \ref{Sect_CR2066}.

\section{Implicit Reconstruction of some Coronal Magnetic Field Structures} 
\subsection{Relationship Between the Coronal Electron Density and Coronal Magnetic Field Structures} 
\label{Sect_MHD}

\begin{figure}[b!]
\centering{
\includegraphics*[bb=96 299 548 604,width=0.83\textwidth]{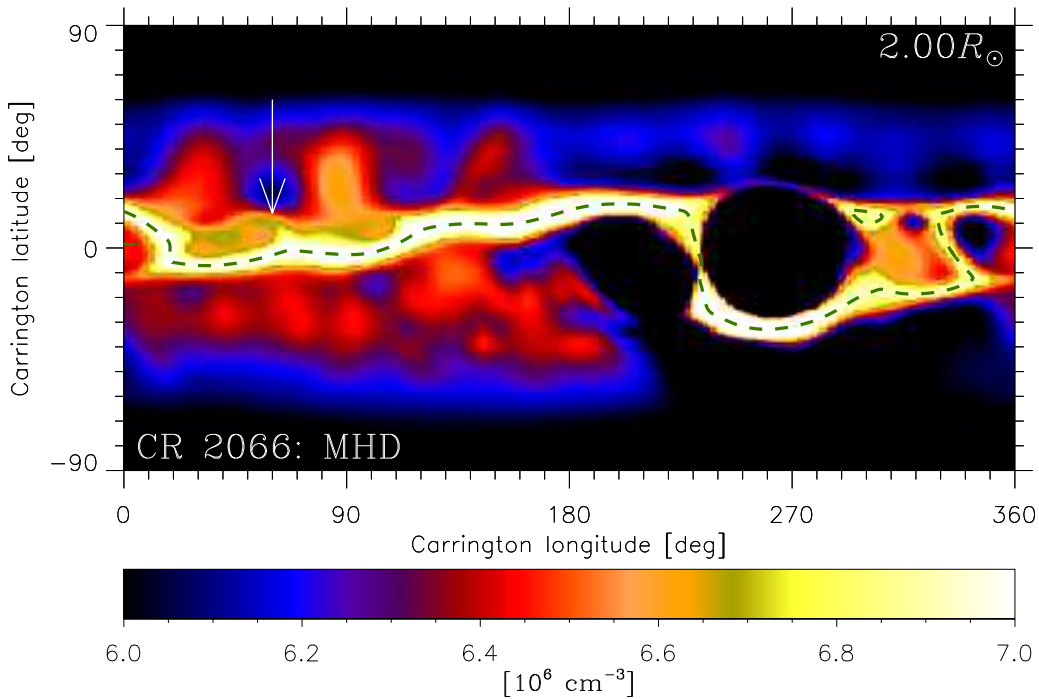}}
\caption{MHD simulation for CR 2066. 
Spherical cross-section of the electron density at heliocentric distance of $2 \ \mathrm{R}_\odot$. 
Dashed line marks position of the magnetic neutral line. }
\label{Fig_MHD_CR2066_sph}
\end{figure}

To establish the relationship between the coronal electron density and the corresponding magnetic field structures, 
we used the results from 3D MHD simulations based on synoptic magnetograms for the specified time period. 
We first investigate the results from a simpler polytropic MHD model \citep{Riley_2001} for CR 2066. 
In this model, the energy equation is simplified by assuming a polytropic equation of state, 
with a reduced polytropic index $\gamma = 1.05$, 
in the spirit of the original model for the solar wind of \cite{Parker_1963}. 
The description of the energy transport in the solar corona provided by the simplified polytropic model is less accurate 
than the full thermodynamic model (see below). 
However, because the polytropic model is computationally more efficient, its solutions can be obtained more routinely. 
It is known that the polytropic model does not estimate the
coronal plasma density and temperature accurately, a result that we confirm and discuss below. 
The group at Predictive Science, Inc., has produced a set of polytropic 
MHD solutions for all of the Carrington rotations in the STEREO era, which are 
available online at \url{www.predsci.com/stereo/}. 
We selected the specific solution for CR 2066, which was based on the SOHO/MDI synoptic magnetic-field data 
measured during the period 15 January -- 21 February 2008. 
The radial component of the magnetic field inferred from the MDI data is used as a boundary condition for the model 
at the lower radial boundary.

\begin{figure}[b!]
\includegraphics*[bb=95 359 547 710,width=\linewidth]{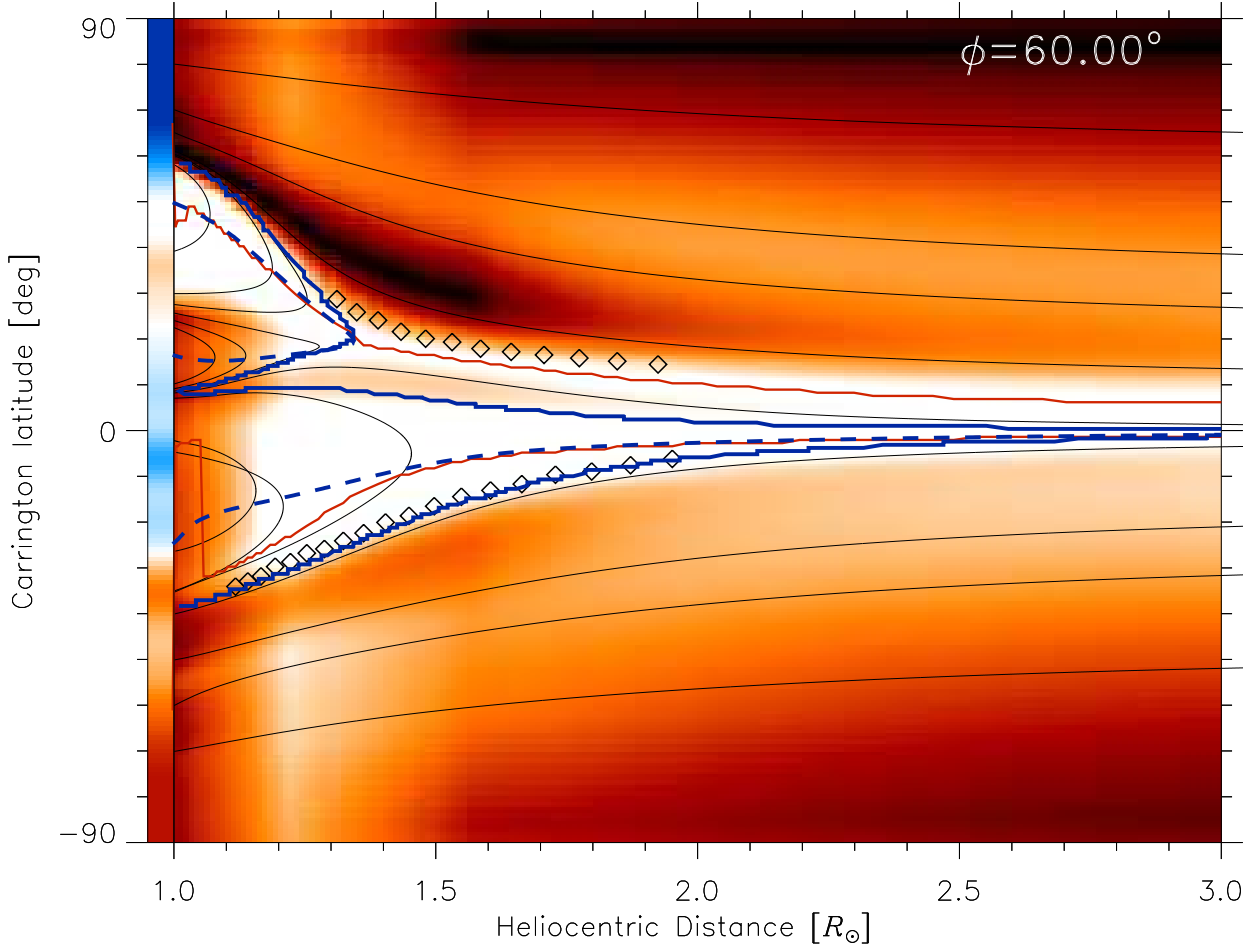}
\caption{MHD simulation for CR 2066. 
Meridional cross-section of the radially filtered and re-scaled electron density at Carrington longitude of $60^\circ$. 
Blue dashed lines mark position of the magnetic neutral line. 
Solid blue lines show position of boundaries between closed and open magnetic-field structures. 
Black diamonds mark the positions of the highest electron-density gradient.
Black lines are magnetic-field lines. 
Red solid lines show the positions of the highest density.
The color bar on the latitude axis shows the value of radial component of the magnetic field [$\vec{B_r}$] 
at the photospheric level, which was used as a lower boundary condition in the simulations.}
\label{Fig_MHD_CR2066_hires60}
\end{figure}

Figures \ref{Fig_MHD_CR2066_sph} and \ref{Fig_MHD_CR2066_hires60} represent result of the polytropic MHD model for CR 2066. 
Figure \ref{Fig_MHD_CR2066_sph} shows the spherical cross-section of the electron density 
at a heliocentric distance of $2 \ \mathrm{R}_\odot$. 
The dashed line marks the magnetic neutral line (where $B_r=0$). 
The density distribution is characterized by two main structures: 
the most dense central structure associated with the magnetic neutral line -- streamer belt, 
and smaller less dense structures connecting neighboring ``peaks'' in latitude direction of the streamer belt. 
The latter, called pseudo-streamers, do not coincide with the magnetic neutral line. 
One of the pseudo-streamers located at Carrington longitude $60^\circ$ is marked by a white arrow in the figure.

Figure \ref{Fig_MHD_CR2066_hires60} shows meridional cross-sections of the electron density for $\phi=60^\circ$. 
In this meridional cross-section, the image for density values was processed through a radial filter 
and re-scaled with a scaling factor depending on the height to magnify low density structures. 
This makes impossible to show the color bar scale for the density in this figure. 
For the streamer region, the highest densities 
represent either the position of the magnetic neutral line (and the current sheet) or 
magnetic-field lines originating from regions with higher electron density following the loop structures. 
For the pseudo-streamer region in the closed field region, 
the behavior of the highest density positions is similar to those for the streamer region. 
For the pseudo-streamer region in the open field region, 
the highest density position follows the behavior of the magnetic-field line. 
Black diamonds in Figure \ref{Fig_MHD_CR2066_hires60} mark the positions of the highest density gradient at fixed heliocentric distances. 
These positions follow the behavior of the magnetic-field lines, 
and for the streamer region they coincide with the boundary position between the closed and open magnetic field structure. 
From this we can deduce a general qualitative picture of the coronal magnetic-field structures directly from the 
reconstructed 3D electron density structure.

\subsection{Retrieving Coronal Magnetic Field Structures from the EUVI 195 \AA \  Emissivity} 
\label{Sect_Test_Tomo_EUVI}

In this subsection we examine the relationship between the location of closed magnetic field regions
and the 3D structure of the EUVI 195 \AA \  emissivity.

\begin{figure}[b!]
\centering{
\includegraphics*[bb=96 359 548 604,width=0.83\textwidth]{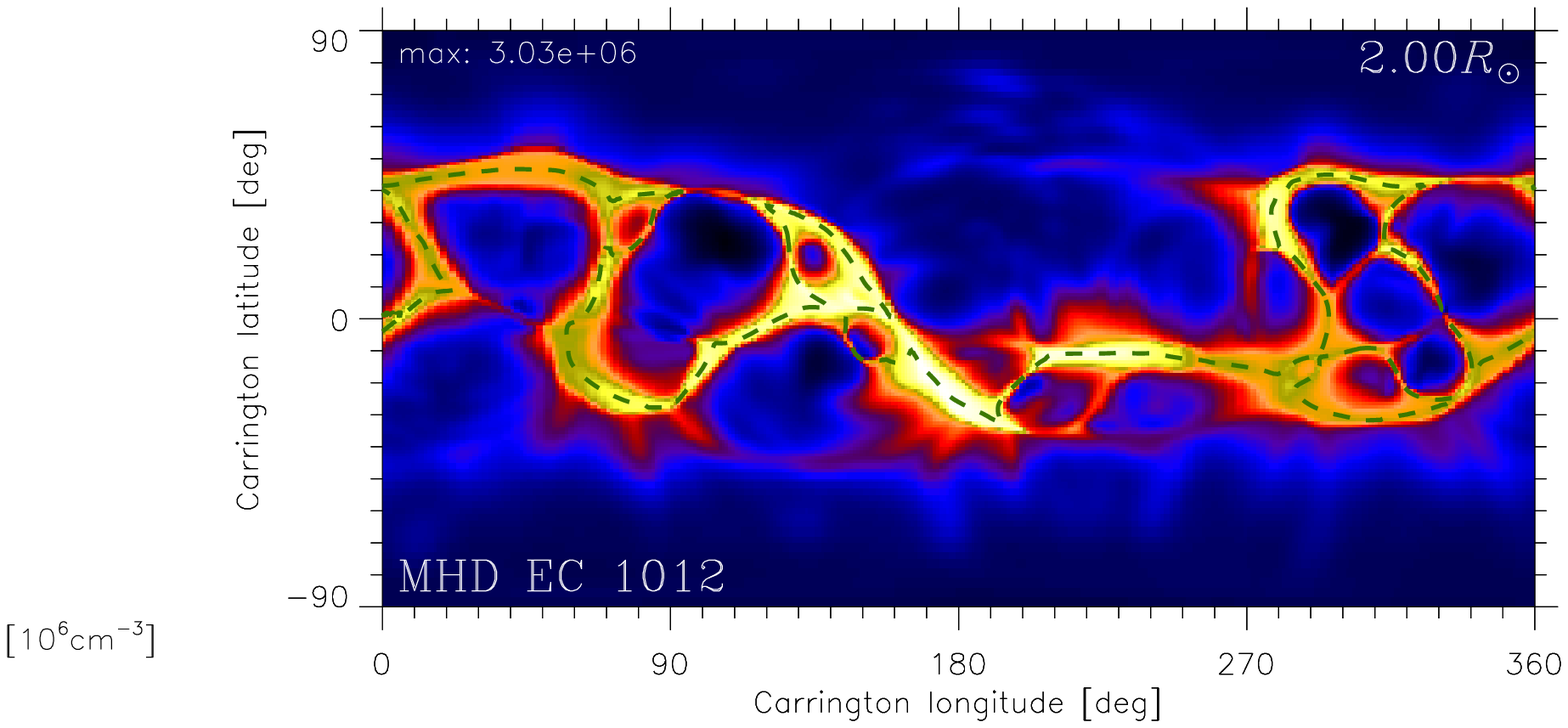}\\
\includegraphics*[bb=96 299 548 359,width=0.83\textwidth]{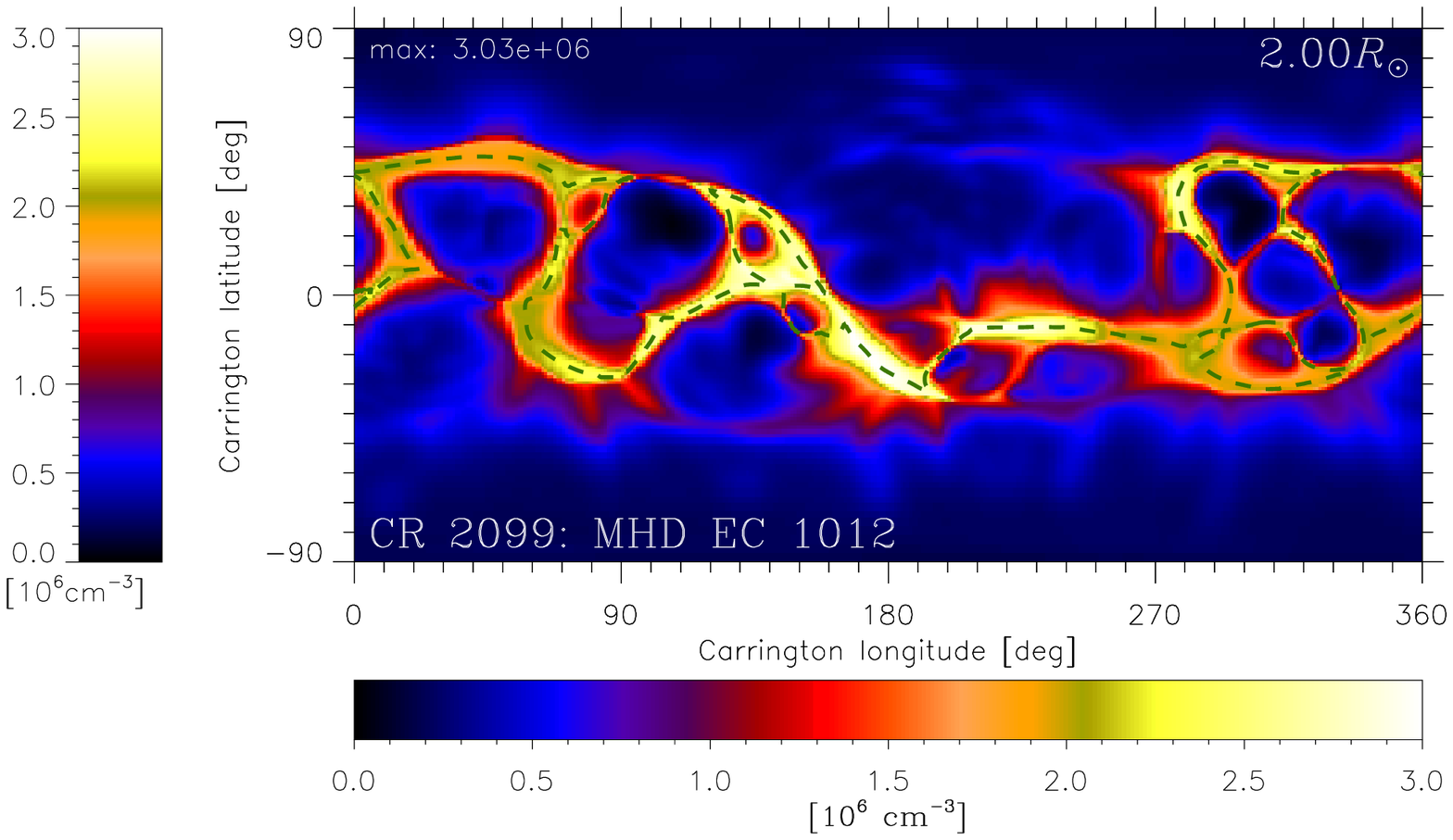}}
\caption{
The electron density at heliocentric distance of $2 \ \mathrm{R}_\odot$ from the thermodynamic MHD simulation for the solar eclipse of 11 July 2010. 
Dashed lines denote the magnetic neutral line. }
\label{Fig_MHD_EC1012_sph}
\end{figure}

\begin{figure}[h!]
\includegraphics*[bb=97 362 353 708,width=0.49\linewidth]{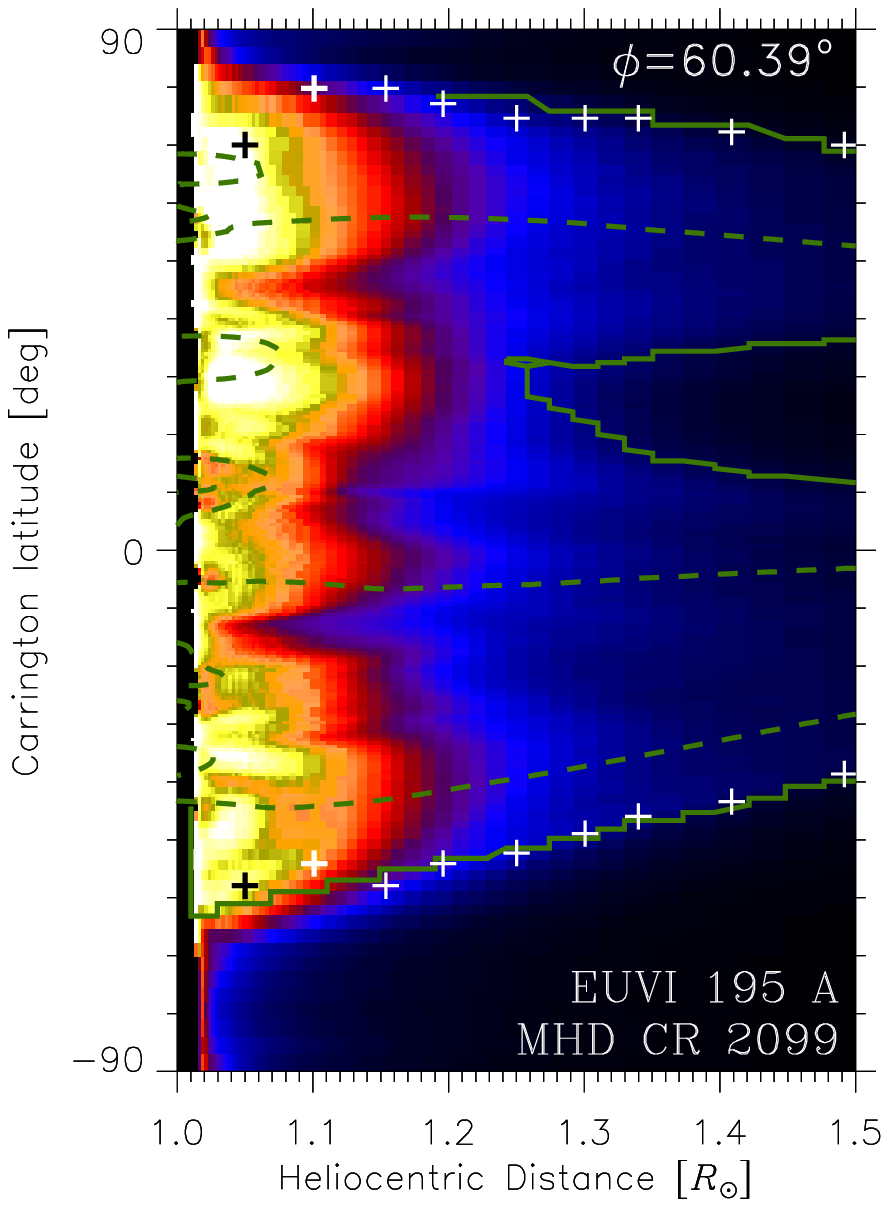}
\put(-65.0,37.0){\tikz \fill [black] (0.0,0.0) rectangle (1.9,0.4);}
\includegraphics*[bb=97 362 353 708,width=0.49\linewidth]{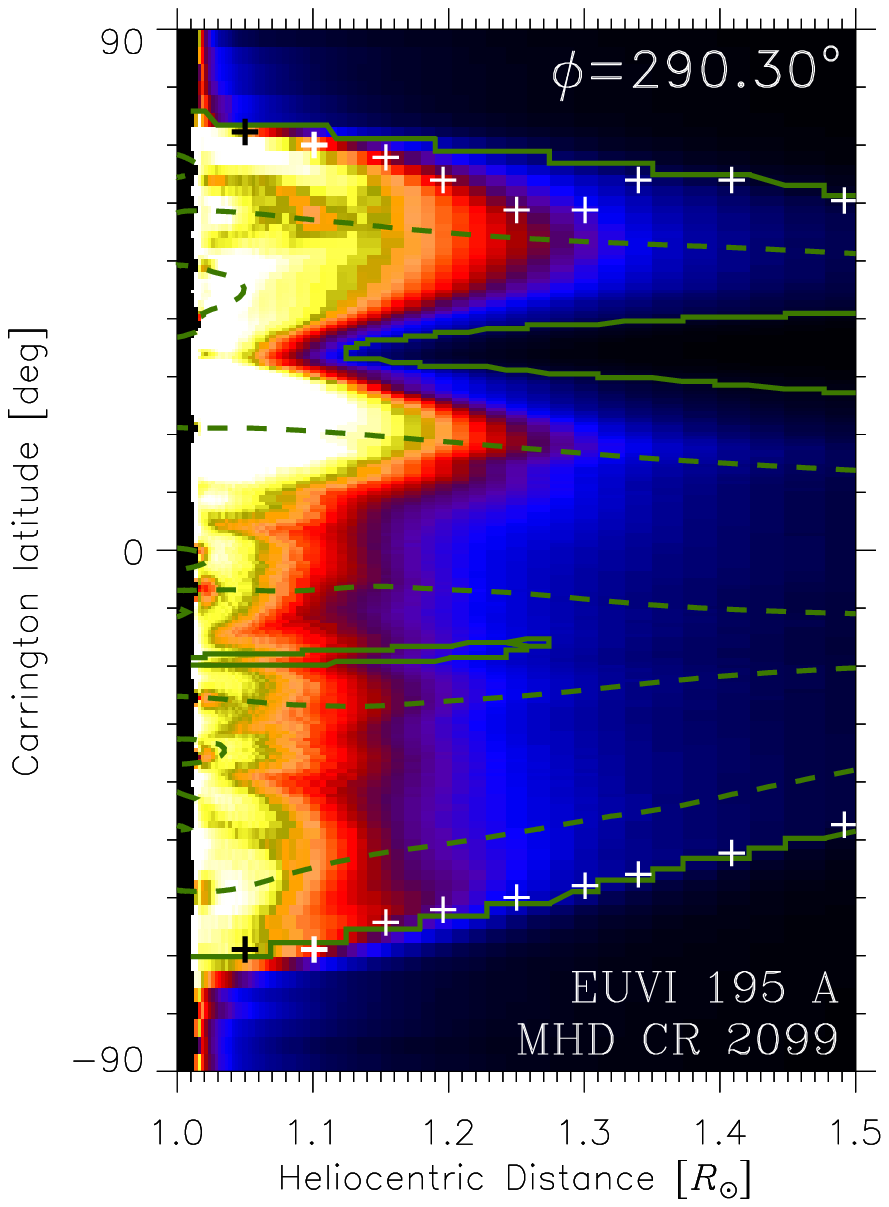}
\put(-65.0,37.0){\tikz \fill [black] (0.0,0.0) rectangle (1.9,0.4);}
\caption{MHD simulation for the 11 July 2010 solar eclipse. 
Meridional cross-section of the EUVI 195 \AA \  emissivity at Carrington longitudes of $60^\circ$ (left) and $290^\circ$ (right). 
Green dashed lines mark position of the magnetic neutral line. 
Solid green lines show the boundary positions between closed and open magnetic-field structures.
Crosses mark the highest-emissivity gradient in the latitudinal direction in the thermodynamic MHD model.
The color bar on the latitude axis shows the radial component of the magnetic field [$\vec{B_r}$] 
at the photospheric level that was used as lower boundary condition in the simulations.}
\label{Fig_EC1012_EUVI_hires_two}
\end{figure}

As mentioned in Section \ref{Sect_MHD}, the polytropic MHD model for CR 2066 predicts a much lower range of density values 
at a fixed heliocentric distance than is observed in reality.  
This is principally due to the overly simplified polytropic energy equation. 
This is a recognized shortcoming of the polytropic MHD model that has been addressed in recent improvements to the model. 
The newer thermodynamic MHD model uses an improved equation for energy transport in the corona 
that includes parallel thermal conduction along the magnetic-field lines, 
radiative losses, and parameterized coronal heating \citep{Lionello_2009}. 
This thermodynamic MHD model produces more accurate estimates of plasma density and temperature in the corona. 
A detailed description is given by \cite{Mikic_2007} and \cite{Lionello_2009}. 
Its application to the total solar eclipse of 1 August 2008 was described by \cite{Rusin_2010}.

The thermodynamic simulation used SOHO/MDI magnetic-field data measured from \mbox{10 June -- 4 July,} 2010 (a combination of CR 2097 and 2098), 
and an extension of the coronal heating model described by \cite{Lionello_2009}. 
The results of this simulation were used to predict the structure of the corona for the solar eclipse of 11 July 2010 
\footnote{\url{www.predsci.com/corona/jul10eclipse/jul10eclipse.html}}. 
This model was also used to produce the artificial data for testing the tomography method for uncertainties, 
as described in Section \ref{Sect_Test_Tomo_COR1}. 
Figure \ref{Fig_MHD_EC1012_sph} shows a spherical cross-section of the electron density 
at a heliocentric distance of $2 \ \mathrm{R}_\odot$.

Figure \ref{Fig_EC1012_EUVI_hires_two} shows meridional cross-sections of the EUVI 195 \AA \  emissivity at the longitudes of $60$ 
and $290^\circ$ for the MHD result of the solar eclipse of 11 July 2010. 
Green dashed lines mark position of the magnetic neutral line. 
Solid green lines show the boundary position between closed and open magnetic-field structures. 
Crosses mark the highest emissivity gradient in latitudinal direction. 
In most cases, the highest emissivity gradient coincide with the 
boundaries between closed and open magnetic-field regions. 
But sometimes these positions are shifted toward regions with higher emissivity, which is indicative of closed magnetic-field regions. 
If we assume that the boundary between open and closed field structures is related to the highest density gradient, 
as shown in Section \ref{Sect_MHD}, Figures \ref{Fig_MHD_CR2066_hires60} and \ref{Fig_EC1012_hires_two}, 
then this shift can be explained as a result of the dependence of the emissivity on the square of the electron density, 
{\it i.e.} $\varepsilon$ varies as $G(T,N_e)N_e^2$, 
where $N_e$ is electron density, $T$ is the electron temperature, and $G(T,N_e)$ is the line contribution function. 
Thus, this dependence could cause a shift of the highest gradient towards a region with higher density values.

\section{Estimation of Uncertainties in the Tomography} 
\label{Sect_Test_Tomo_COR1}

As described in Section \ref{Sect_Tomo}, 
the 3D coronal density was obtained by the regularized tomographic inversion, 
where we used the smoothing operation as a regularization. 
This method can introduce a small systematic error in the reconstructed density values which, in turn, 
could generate the errors in the locations of the highest density gradient. 
Therefore, to estimate this systematic error, 
we used the 3D MHD model to create artificial pB-data for the tomographic inversion with the same temporal (or angular) and 
spatial sampling as in real data 
(28 images during half of a solar rotation, where each $340\times 340$ image covers the FOV with radius of $4\ \mathrm{R}_\odot$). 

\begin{figure}[p]
\centering{
\includegraphics*[bb=95 377 547 710,width=0.7\linewidth]{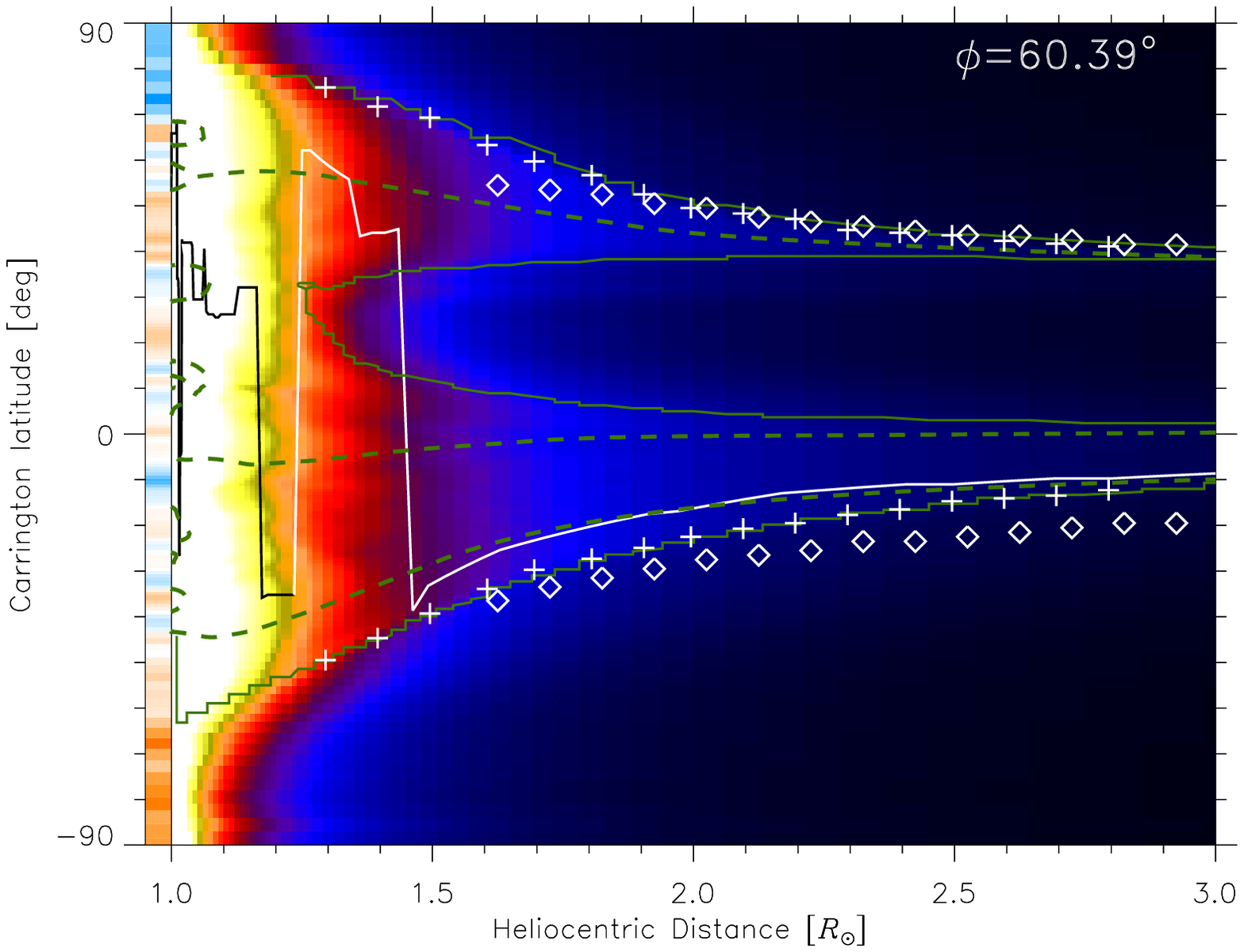} \\
\includegraphics*[bb=95 359 547 710,width=0.7\linewidth]{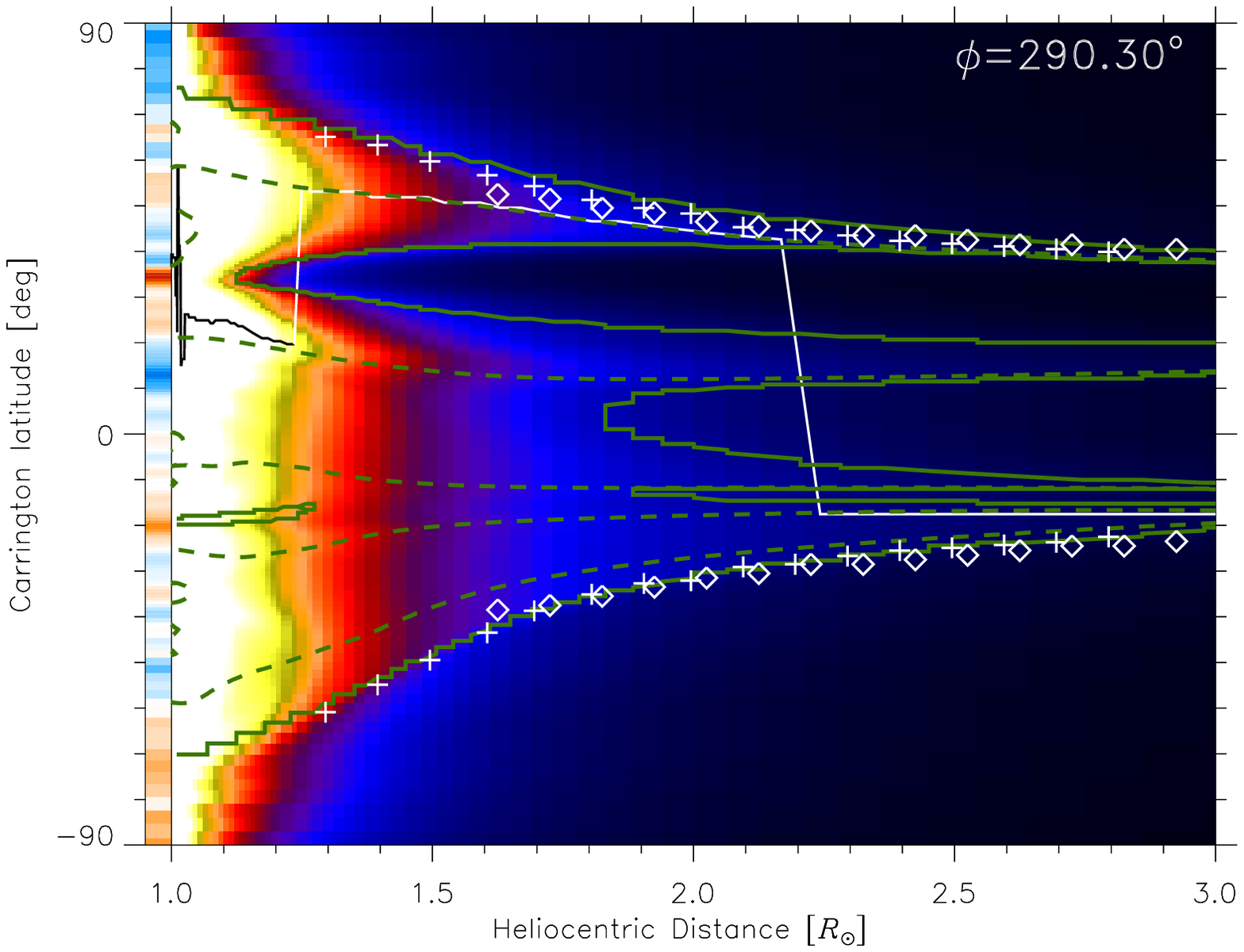}}
\caption{MHD simulation for 11 July 2010 solar eclipse. 
Meridional cross-section of the electron density at Carrington longitudes of $60^\circ$ (top panel) and $290^\circ$ (bottom panel). 
Green dashed lines denote position of the magnetic neutral line. 
Solid green lines show the boundaries between closed and open magnetic-field structures. 
Crosses and diamonds highlight the highest electron-density gradient for the results of the thermodynamic 
MHD model and the test tomographic reconstruction based on simulated COR-1 data, respectively. }
\label{Fig_EC1012_hires_two}
\end{figure}

We used the thermodynamic MHD model described in Section \ref{Sect_Test_Tomo_EUVI} 
to produce the artificial pB-data and test the tomography method for errors. 
Figure \ref{Fig_EC1012_hires_two} shows meridional cross-sections of the electron density at longitudes of $60$ and $290^\circ$ 
for the MHD result of the 2010 solar eclipse. 
Green dashed lines mark position of the magnetic neutral line. 
Solid green lines show the boundaries between closed and open magnetic-field structures. 
White crosses and diamonds highlight the highest density gradients at fixed heliocentric distances 
for the MHD model and tomography results, respectively. 
The error in determining the highest density gradient does not exceed $10^\circ$ 
and the reconstructed positions tend to be less curved at heliocentric distances near the lower limit of 
the reconstruction domain ($1.5\ \mathrm{R}_\odot$). 
This demonstrates that the highest density gradient obtained by the tomography 
can be used to determine the boundaries between closed and open magnetic-field structures.

\section{3D Coronal Structure during CR 2066}
\label{Sect_CR2066}

\begin{figure}[t!]
\centering{
\includegraphics*[bb=96 299 555 604, width=0.7\linewidth]{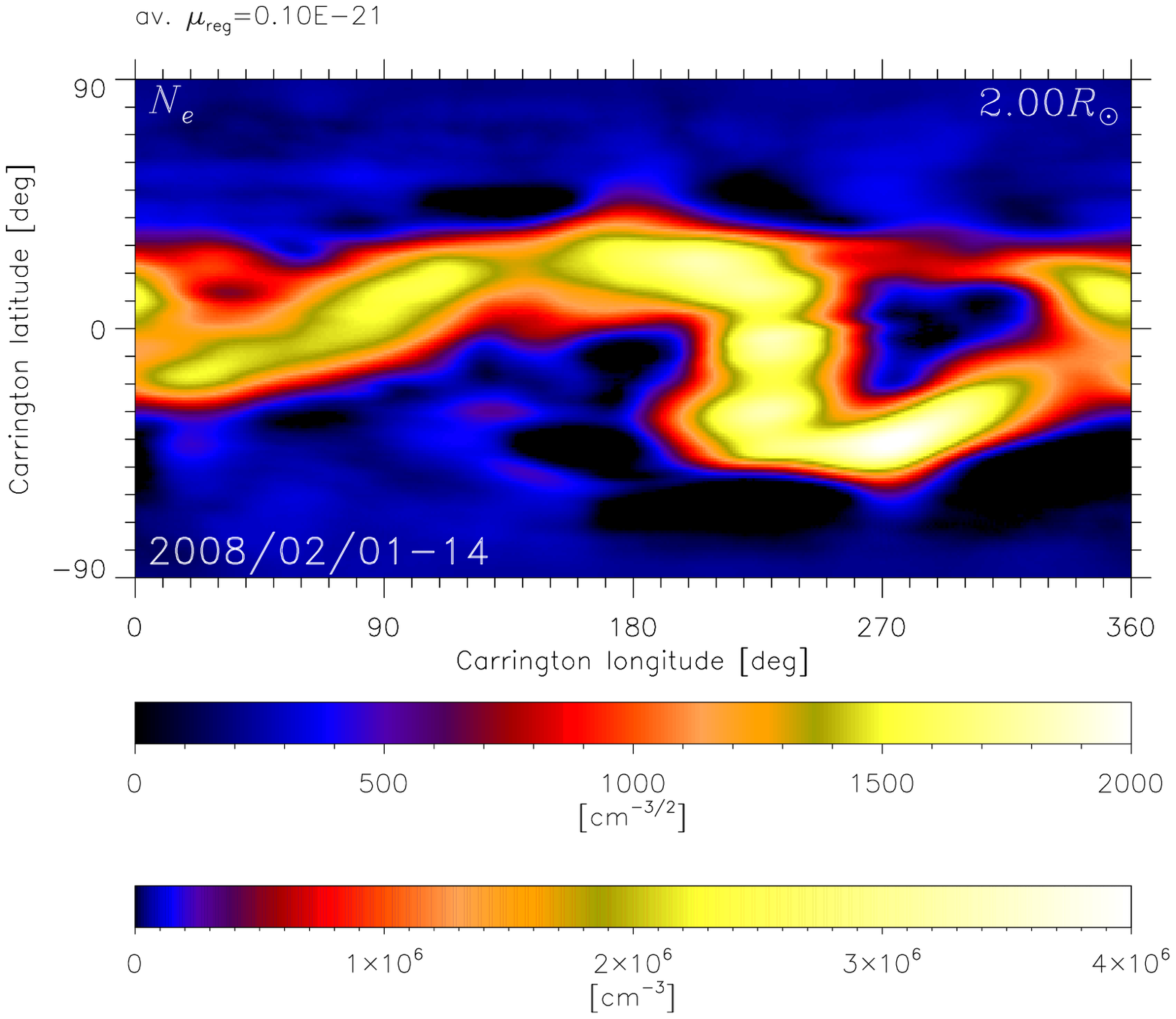}}
\caption{Spherical cross-section of the reconstructed electron density in square-root scale at heliocentric distance of $2 \ \mathrm{R}_\odot$. 
The reconstruction is obtained by tomography based on COR-1 data obtained during February 1--14, 2008 (CR 2066).}
\label{Fig_CR2066_Rec_sph}
\vspace{0.7cm}
\centering{
\includegraphics*[bb=64 359 548 627, width=0.7\linewidth]{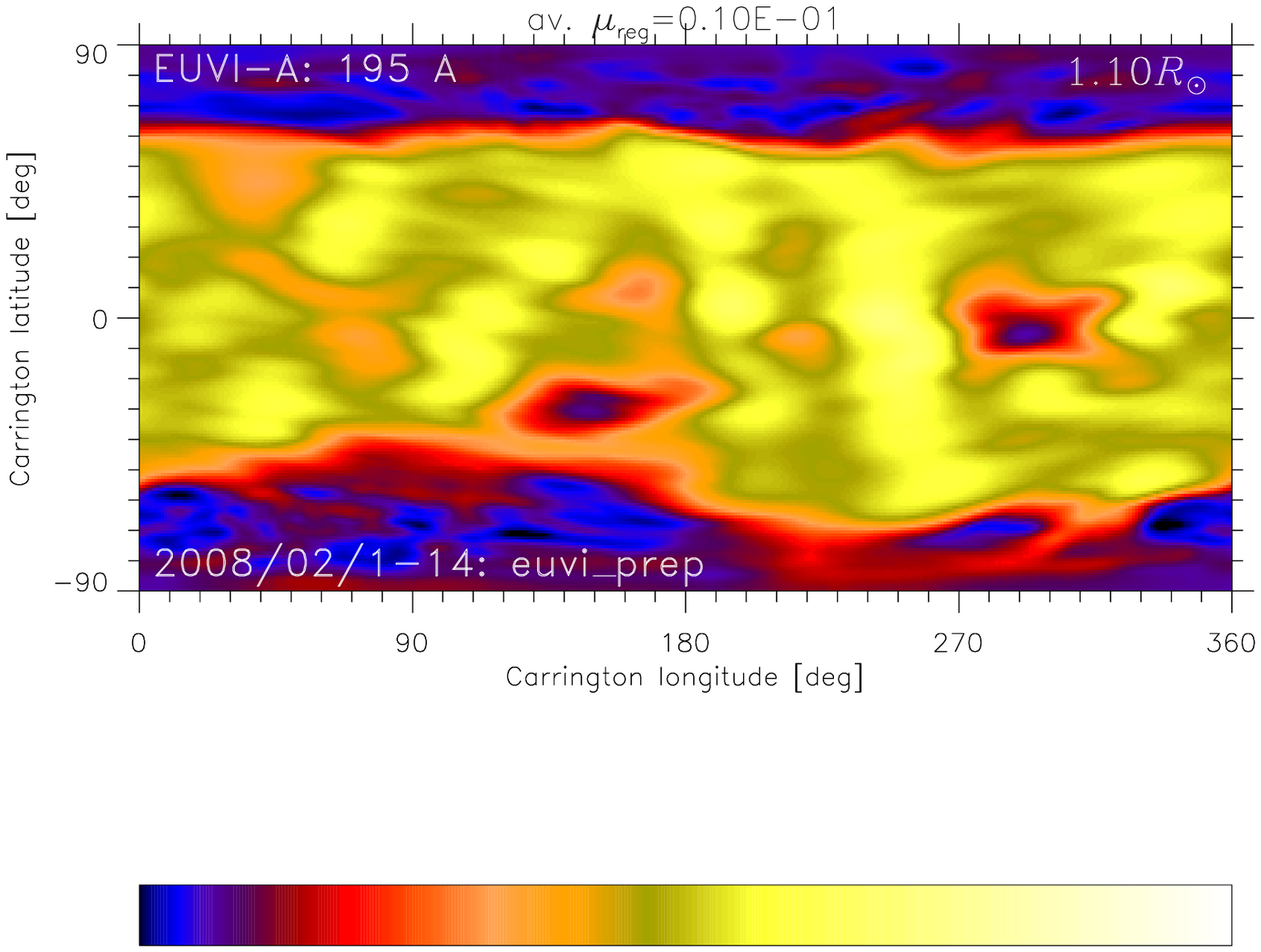}\\
\includegraphics*[bb=64 299 548 359, width=0.7\linewidth]{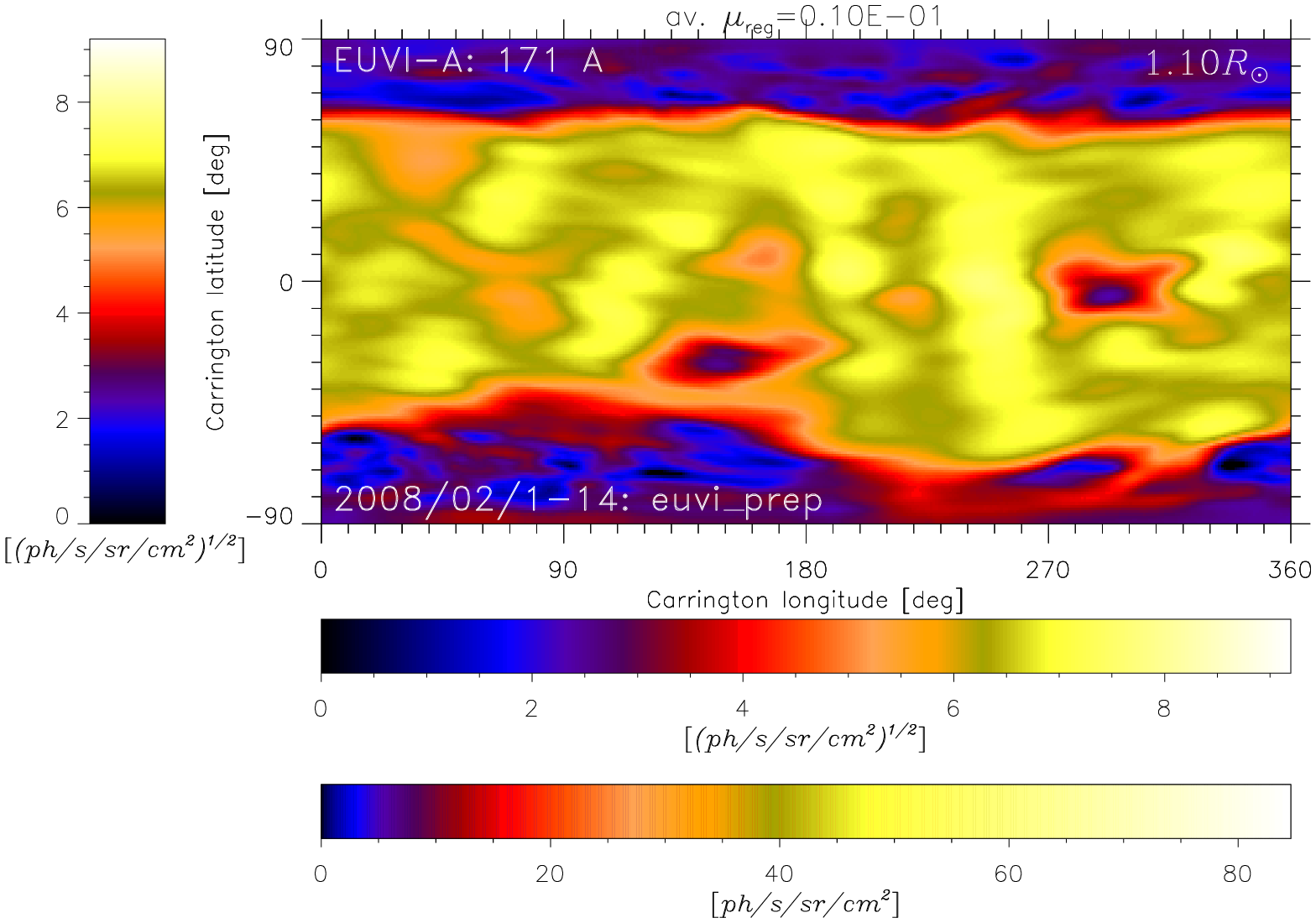}}
\caption{Spherical cross-section of the reconstructed 3D EUVI 195 \AA \  emissivity in square-root scale 
at heliocentric distances of $1.1 \ \mathrm{R}_\odot$. 
The reconstruction is obtained by tomography based on EUVI data obtained during 1--14 February 2008 (CR 2066).}
\label{Fig_CR2066_EUVI_Rec_sph}
\end{figure}

CR 2066 represents the deep minimum of the solar-activity cycle. 
Therefore, the 3D corona during CR 2066 is ideally suited for studying with the tomography method 
because the reconstruction errors are minimized due to low coronal activity. 
We performed two types of tomographic reconstructions: 
a 3D reconstruction for the electron density based on STEREO/COR1 data, 
and a 3D reconstruction for the EUVI 195 \AA \  emissivity [units of photons s$^{-1}$ sr$^{-1}$ cm$^{-3}$] based on STEREO/EUVI data. 
To demonstrate the general structure of the coronal streamer belt for CR 2066, 
Figure \ref{Fig_CR2066_Rec_sph} shows a spherical cross-section of the electron density at 
a heliocentric distance of $2 \ \mathrm{R}_\odot$, 
and Figure \ref{Fig_CR2066_EUVI_Rec_sph} shows the spherical cross-section of the EUVI 195 \AA \  emissivity at 
a heliocentric distance of $1.1 \ \mathrm{R}_\odot$.

\begin{figure}[p]
\renewcommand{\baselinestretch}{1.0} 
\vspace{-0.5cm}
\hspace*{\fill}\includegraphics*[bb=150 410 547 705,width=0.43\linewidth]{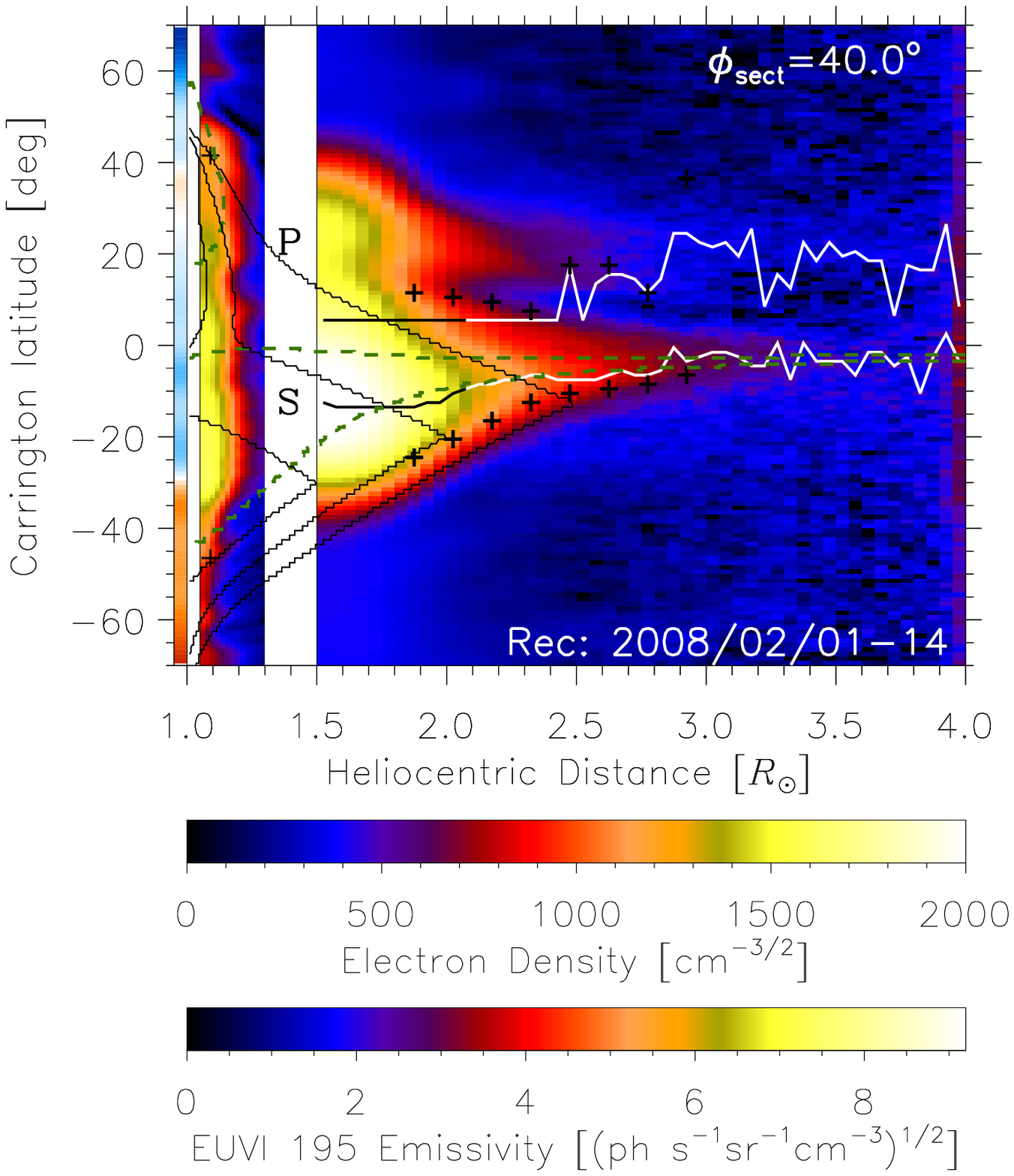}
\hspace*{\fill}\includegraphics*[bb=150 410 547 705,width=0.43\linewidth]{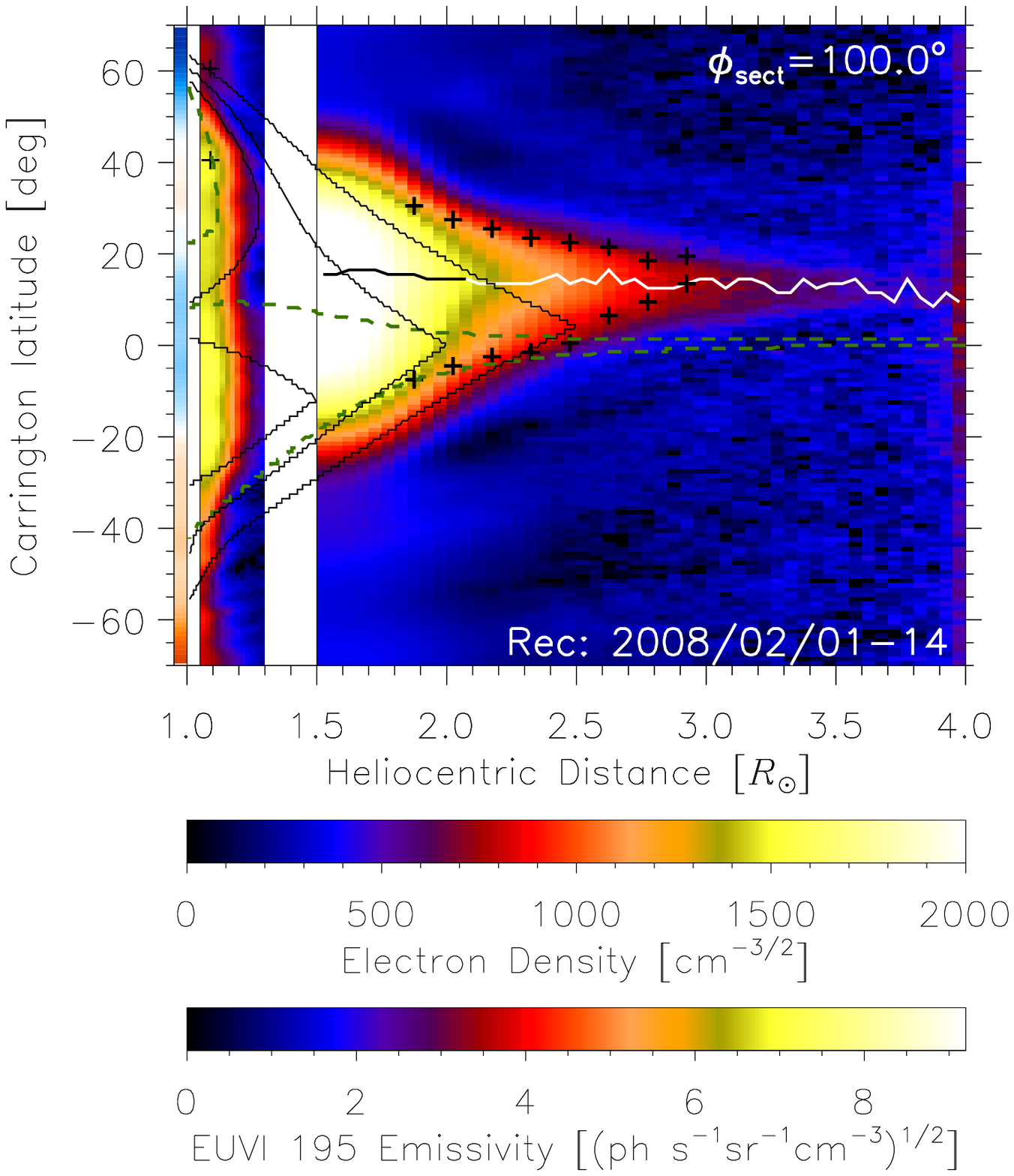}\hspace*{\fill}\\
\hspace*{\fill}\includegraphics*[bb=150 410 547 705,width=0.43\linewidth]{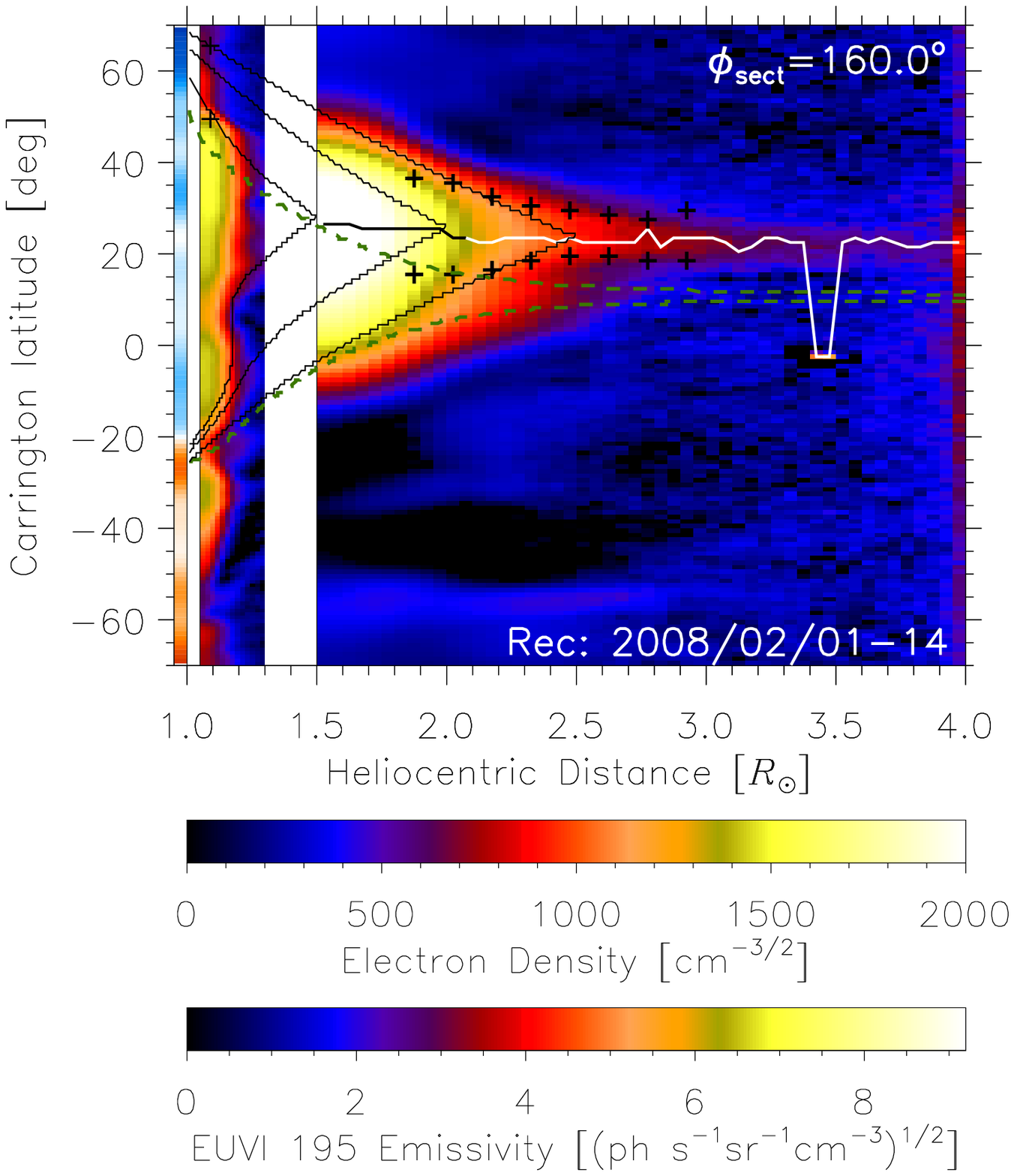}
\hspace*{\fill}\includegraphics*[bb=150 410 547 705,width=0.43\linewidth]{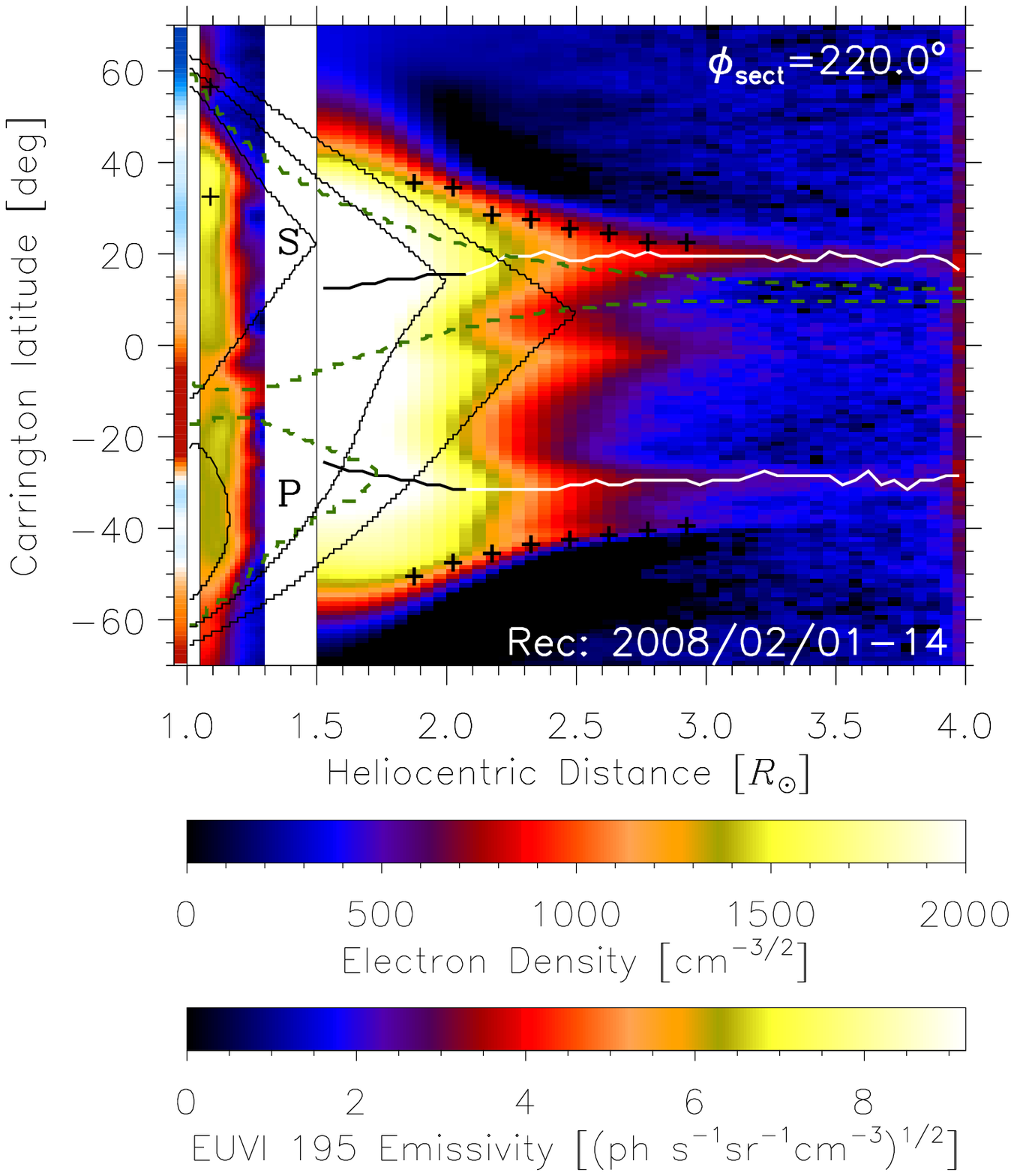}\hspace*{\fill}\\
\hspace*{\fill}\includegraphics*[bb=150 359 547 705,width=0.43\linewidth]{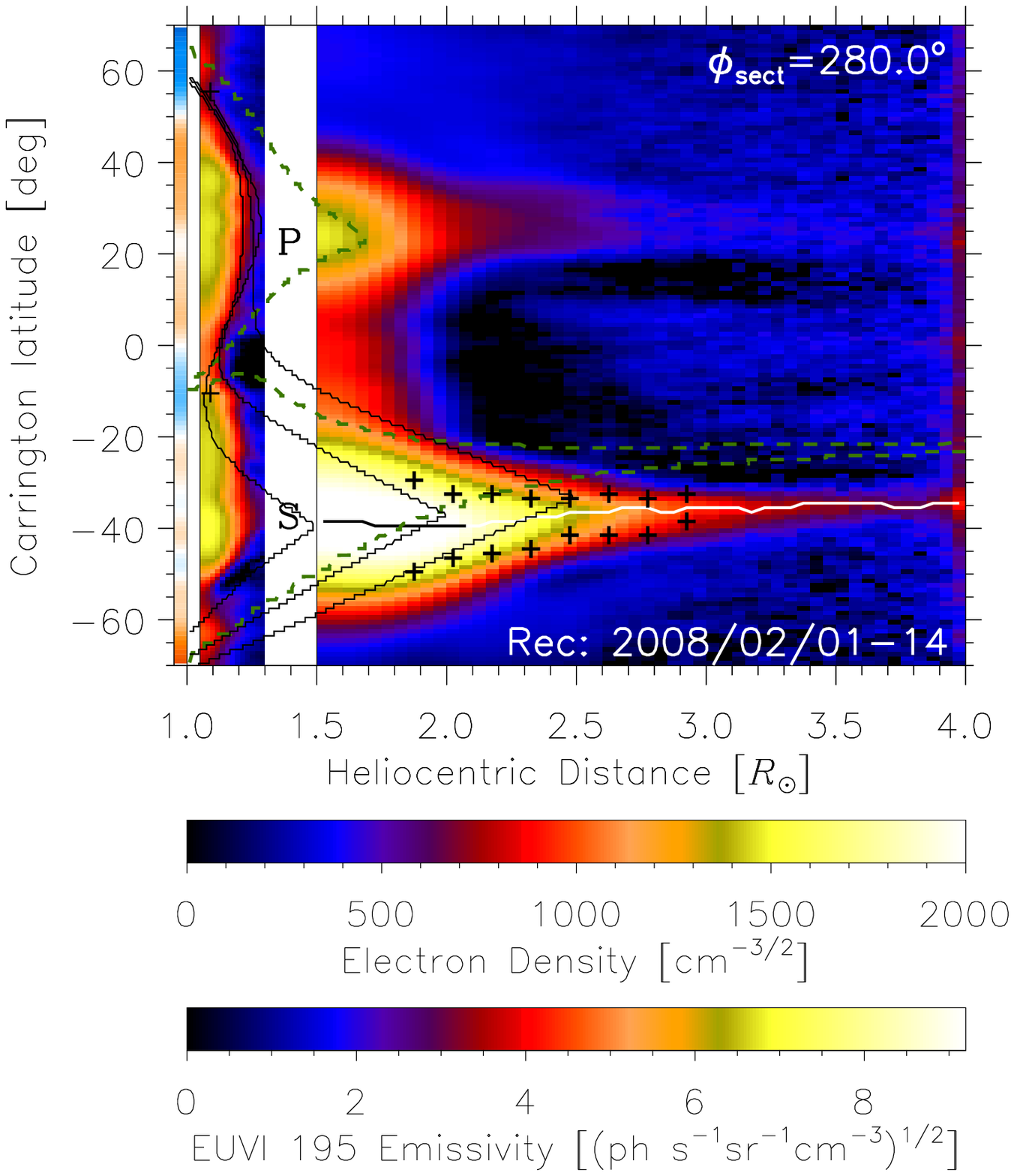}
\hspace*{\fill}\includegraphics*[bb=150 359 547 705,width=0.43\linewidth]{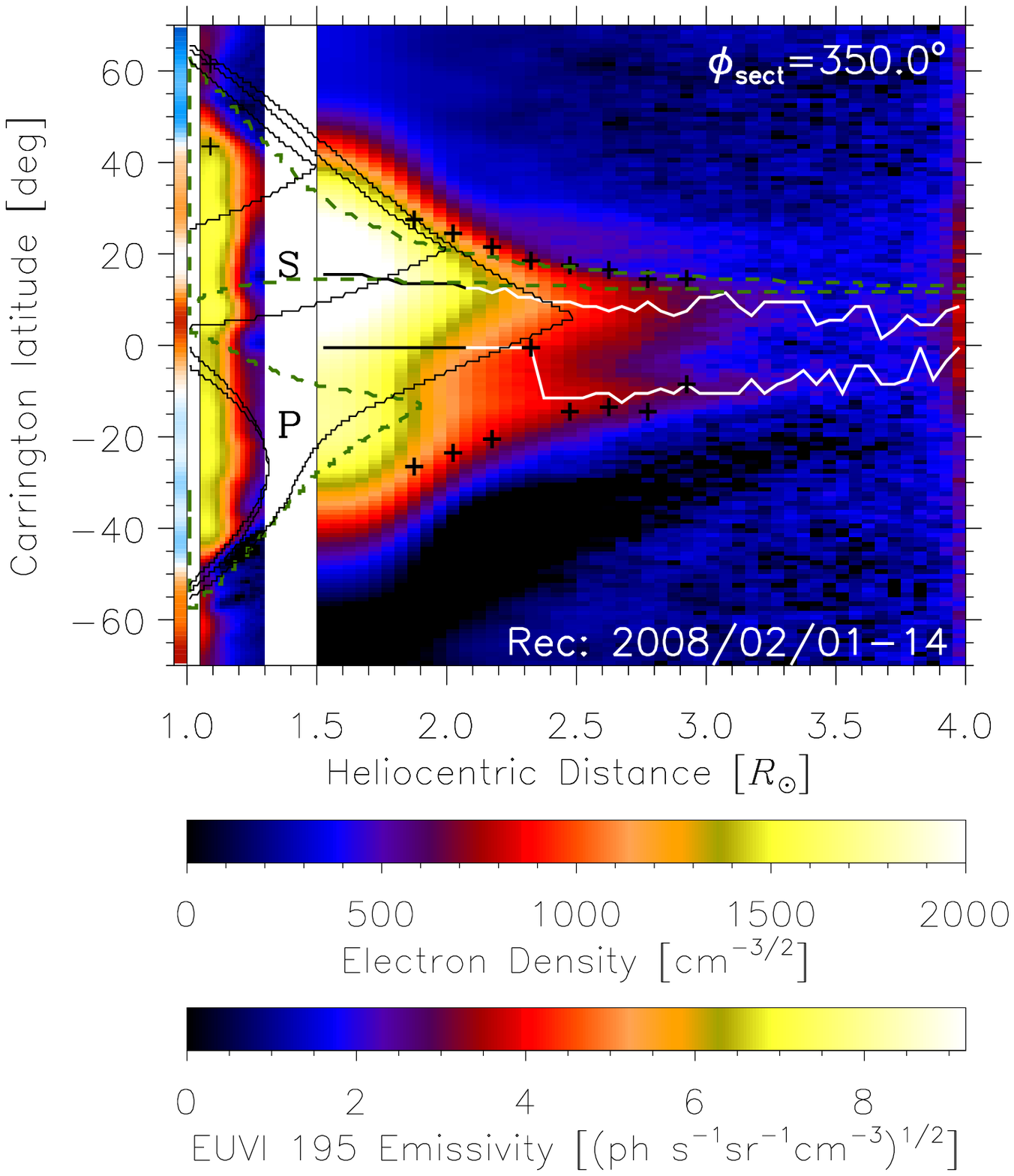}\hspace*{\fill}\\
\hspace*{\fill}\includegraphics*[bb=150 285 547 357,width=0.43\linewidth]{f8f.ps}
\hspace*{\fill}\includegraphics*[bb=150 207 547 279,width=0.43\linewidth]{f8f.ps}\hspace*{\fill}
\vspace{-0.3cm}
\caption{\small{
Reconstructions for CR 2066 based on COR1 data (electron density in the range from $1.5$ to $4 \ \mathrm{R}_\odot$) 
and EUVI 195 \AA \  data (emissivity in the range from $1.05$ to $1.29 \ \mathrm{R}_\odot$). 
Cross-sections for Carrington longitudes of $40$, $100$, $160$, $220$, $280$, and $350^\circ$ are shown from left to right and upper to lower panels, respectively. 
Black and white lines mark the highest density. 
The contour black lines show boundaries between open and closed magnetic-field structures for 
the PFSS models with the source surface located at $1.5$, $2.0$, and $2.5 \ \mathrm{R}_\odot$. 
The dashed green lines show boundaries between open and closed magnetic-field structures for the polytropic MHD model. 
Crosses mark the highest density gradient at several heliocentric distances. 
The color bar on the left side of the latitude axis is corresponding meridional cross-sections 
for the radial component of the photospheric magnetic field used as a boundary condition for the MHD simulation. 
Letters ``S'' and ``P'' mark streamers and pseudo-streamers, respectively. }}
\label{Fig_RSS_COR1_phi_CR2066}
\end{figure}

Figure \ref{Fig_RSS_COR1_phi_CR2066} shows several meridional cross-sections of the electron density 
(range from $1.5$ to $4 \ \mathrm{R}_\odot$) 
and EUVI 195 emissivity (range from $1.05$ to $1.29 \ \mathrm{R}_\odot$). 
A figure with a set of all cross-sections is available in the electronic suplemental material. 
The superimposed black--white lines plotted on the cross-sections shows the highest density. 
Therefore, the lines represent either the magnetic neutral line (and the current sheet), 
or magnetic field lines originating from regions with higher electron density (see Section \ref{Sect_Test_Tomo_COR1} for justification). 
In most of the cross-sections, the superimposed lines become asymptotically radial at about $3 \ \mathrm{R}_\odot$.

Black crosses in Figure \ref{Fig_RSS_COR1_phi_CR2066} show the highest density gradient 
at several heliocentric distances. 
Sometimes they are scattered over a wide range in latitude because of the reconstruction errors, 
which are most probably caused by coronal dynamics and/or noises in input data, 
but in most cases (for example at longitudes of 10, 50, 60, 90, 100, 160, 190, 350$^\circ$) 
they smoothly follow and specify either the magnetic field lines or the boundary between closed and open magnetic-field structures. 
At a heliocentric distance of $1.1 \ \mathrm{R}_\odot$, 
the black crosses mark the highest EUVI 195 emissivity gradient. 
The highest density gradients for tomographic reconstruction based on COR1 data are consistent with 
the highest emissivity gradient for tomographic reconstruction of EUVI 195 \AA \  emissivity. 
Therefore, the locations of the black crosses suggest that the coronal magnetic field near the streamer belt 
becomes radial at about $2.5 \ \mathrm{R}_\odot$ and higher. 

The green dashed contour lines in Figure \ref{Fig_RSS_COR1_phi_CR2066} show the boundaries 
between open and closed magnetic field structures in the polytropic MHD model. 
Although the boundaries between open and closed magnetic field structures in the polytropic MHD model 
do not fully coincide with those derived by the tomography, 
the magnetic field lines in the MHD model near the current sheet positions become asymptotically radial 
at $\approx 2.5 \ \mathrm{R}_\odot$. 
Moreover, the MHD model provides a guidance about how to distinguish 
which coronal structures in the tomographic 3D reconstructions are correlated to streamers, as opposed to pseudo-streamers: 
the ``legs'' of the boundary lines are rooted at the photospheric level in magnetic fields of opposite polarity in the case of streamers,
and in fields of the same polarity in the case of pseudo-streamers. 
Since the latest thermodynamic MHD model produces more accurate estimates of the coronal plasma density and temperature, 
a more detailed comparison between coronal densities estimated from MHD models and observations will be performed in future work.

The black contour lines in Figure \ref{Fig_RSS_COR1_phi_CR2066} show boundaries between open and closed magnetic field structures 
in three PFSS models with Source Surface heliocentric distances [$R_\textrm{ss}$] at $1.5$, $2.0$, and $2.5 \ \mathrm{R}_\odot$. 
The PFSS model with $R_\textrm{ss}=2.5 \ \mathrm{R}_\odot$ does not coincide with the derived positions of the 
streamer and pseudo-streamer nor with the coronal hole positions indicated by the STEREO/EUVI 195 \AA \  emissivity 3D reconstruction. 
The PFSS model with $R_\textrm{ss}=1.5 \ \mathrm{R}_\odot$ appears to fit the latter two structure types better, 
but does not satisfy the requirement for the field to become radial at about $2.5 \ \mathrm{R}_\odot$. 
Thus, the assumption of the potential nature of the coronal magnetic field is not satisfied even during the deep solar minimum.

\section{Conclusion and Outlook}

We applied STEREO-B/COR1 data for CR 2066 to derive the 3D coronal electron density during the deep solar minimum in February 2008 
using the tomography method. 
We then complemented the density reconstruction with the results of 3D MHD simulations 
and 3D EUVI 195 \AA \  emissivity to determine the relationship between the density, emissivity and magnetic-field structures. 
Specifically, we found that 
\begin{itemize}\itemsep0em
\item the locations of density maximum in the 3D reconstructions can serve as 
an indicator for current sheet and pseudo-streamer positions; 
\item the locations of the highest density gradient in the 3D reconstructions can serve as 
an indicator for boundaries between closed and open magnetic-field structures. 
\end{itemize}
Thus, we showed that 3D coronal electron density reconstruction, 
especially when used in conjunction with 3D EUVI 195 \AA \ emissivity reconstruction, 
and with the guidance provided by state-of-the-art 3D MHD simulations, 
can be instrumental in retrieving the geometry of the global solar coronal magnetic field. 
Specifically, this method can derive 
the locations of boundaries between open and closed magnetic-field structures, 
and distances where the magnetic-field lines become radially directed. 
The nearly realistic 3D coronal electron density and 3D EUVI 195 \AA \ emissivity are both obtained by the tomography method. 
To estimate the error in determining these positions by the tomography, 
we tested the tomographic method with simulated pB-data produced by LOS-integrating the result of the thermodynamic MHD model. 
As a result of this test, we found that the tomography can reliably determine these positions.

We then reconstructed the 3D coronal electron density and EUVI 195 \AA \ emissivity based on real 
STEREO/COR1 and STEREO/EUVI observations, respectively, for CR 2066, which corresponds to deep solar minimum. 
The reconstructed radial dependence of the latitude positions of the highest density and emissivity and its gradient 
suggests that the magnetic field lines become radial at about $2.5 \ \mathrm{R}_\odot$ and higher for most of the longitudinal positions. 
Moreover, we determined the boundaries between regions with open and closed magnetic field structures. 
Because the 3D reconstructions are entirely based on {\it coronal} observations, 
the results can serve as a test and/or as an additional constraint for coronal models. 
As an initial step toward this goal, we analyzed the consistency of the PFSS model for different source surface distances with 
the reconstructed 3D electron density and EUVI 195 emissivity structures. 
We conclude that the assumption of the potential nature of the coronal global magnetic field is 
not satisfied even during the deep solar minimum. 
Would a linear force-free or NLFFF approximation offer a better description of the solar coronal magnetic field? 
How complex do the coronal structures appear during a solar maximum? 
These are topics of a study to be performed in the near future.

\section*{Acknowledgments}

\begin{flushleft}
The final publication is available at Springer via \url{http://dx.doi.org/10.1007/s11207-014-0525-7}
\end{flushleft}


\bibliographystyle{apj}

\tracingmacros=2
\bibliography{biblio}

\end{document}